%% file: main.tex
\documentclass[sigconf,screen]{acmart}
\renewcommand\footnotetextcopyrightpermission[1]{}
\input{__packages}
\input{__macros}

\begin{document}

\title{Harnessing Large Language Models for Seed Generation in Greybox Fuzzing}

\author{Wenxuan Shi}
\email{wenxuan.shi@northwestern.edu}
\affiliation{
  \institution{Northwestern University}
  \country{}
}
\authornote{These authors contributed equally to this work.}

\author{Yunhang Zhang}
\email{u1399304@utah.edu}
\affiliation{
  \institution{University of Utah}
  \country{}
}
\authornotemark[1]

\author{Xinyu Xing}
\email{xinyu.xing@northwestern.edu}
\affiliation{
  \institution{Northwestern University}
  \country{}
}

\author{Jun Xu}
\email{junxzm@cs.utah.edu}
\affiliation{
  \institution{University of Utah}
  \country{}
}

\input{00_abstract}

\maketitle
\fancyhead[L]{}
\fancyhead[C]{}
\fancyhead[R]{}




\input{01_introduction}

\input{02_background}

\input{03_challenges}

\input{04_design}

\input{05_evaluation}

\input{06_relatedwork}

\input{07_conclusion}

\input{appendix}

\clearpage
\bibliographystyle{ACM-Reference-Format}
\bibliography{fse}

\end{document}

%% file: __packages.tex
\usepackage{pifont} 
\usepackage{enumitem} 
\usepackage{xspace} 
\usepackage{tcolorbox} 

\usepackage{amsmath} 
\usepackage{listings} 
\usepackage{xcolor} 

%% file: __macros.tex
\newcommand{\libfuzzer}{{\sc libFuzzer}\xspace}
\newcommand{\magma}{{\sc MAGMA}\xspace}
\newcommand{\sys}{{\sc SeedMind}\xspace}
\newcommand{\ossai}{{\sc OSSFuzz-AI}\xspace}

\newcommand{\afl}{{\sc AFL}\xspace}
\newcommand{\aflpp}{{\sc AFL++}\xspace}
\newcommand{\honggfuzz}{{\sc Honggfuzz}\xspace}

\newcommand{\gptfo}{GPT-4o\xspace}
\newcommand{\gptturbo}{GPT-3.5-Turbo\xspace}
\newcommand{\claudesonnet}{Claude-3.5-Sonnet\xspace}

\newcommand{\ossfuzzprog}{166\xspace}
\newcommand{\ossfuzzharness}{674\xspace}

\newcommand{\code}[1]{{\fontfamily{lmtt}\selectfont{#1}}}

\definecolor{redbar}{RGB}{190,68,61}
\definecolor{blueline}{RGB}{45, 144, 220}
\definecolor{cadmiumgreen}{rgb}{0.0, 0.42, 0.24}
\definecolor{tablegray}{rgb}{0.83, 0.83, 0.83}
\definecolor{light-gray}{gray}{0.95}
\definecolor{shiningblue}{RGB}{0, 150, 255}

%% file: 00_abstract.tex
\begin{abstract}
Greybox fuzzing has emerged as a preferred technique for discovering software bugs, striking a balance between efficiency and depth of exploration.
While research has focused on improving fuzzing techniques, the importance of high-quality initial seeds remains critical yet often overlooked.
Existing methods for seed generation are limited, especially for programs with non-standard or custom input formats.
Large Language Models (LLMs) has revolutionized numerous domains, showcasing unprecedented capabilities in understanding and generating complex patterns across various fields of knowledge.
This paper introduces \sys, a novel system that leverages LLMs to boost greybox fuzzing through intelligent seed generation.
Unlike previous approaches, \sys employs LLMs to create test case generators rather than directly producing test cases.
Our approach implements an iterative, feedback-driven process that guides the LLM to progressively refine test case generation, aiming for increased code coverage depth and breadth.
In developing \sys, we addressed key challenges including input format limitations, context window constraints, and ensuring consistent, progress-aware behavior. Intensive evaluations with real-world applications show that \sys effectively harnesses LLMs to generate high-quality test cases and facilitate fuzzing in bug finding, presenting utility comparable to human-created seeds and significantly outperforming the existing LLM-based solutions. 
\end{abstract}

%% file: 01_introduction.tex
\section{Introduction\label{sec:intro}}

To discover bugs in code, fuzzing~\cite{manes2019art} has been considered one of the most practical techniques, thanks to its easy application to production-grade software. After decades of development, greybox fuzzing~\cite{li2018fuzzing,manes2018art,godefroid2020fuzzing} has emerged as the most preferable option. By leveraging lightweight instrumentation to obtain runtime feedback for driving the exploration, greybox fuzzing strikes a balance between the efficiency of blackbox fuzzing~\cite{godefroid2007random} and the depth of whitebox fuzzing~\cite{godefroid2012sage}. The arising of mature greybox fuzzers like \afl~\cite{fioraldi2023dissecting}, AFL++~\cite{fioraldi2020afl++}, \honggfuzz~\cite{honggfuzz} is furthering the popularity.

Technically, greybox fuzzing starts with a set of initial test cases, called \textit{seeds}, and iteratively derives new test cases from the seeds to expand code coverage. Thus, to achieve better results, \textit{it is critical to prepare a set of high-quality seeds to bootstrap the fuzzing process}. For example, a seed already reaching the buggy site can significantly reduce the difficulty and time cost for fuzzing to trigger the bug. Yet, obtaining good seeds has not been easy.

In practice, the most common strategy to prepare seeds is through manual inspection of the program logic and construction of desired test cases. This strategy might work for programs taking input of popular formats (e.g., XML, HTML, PDF, etc.), as their test cases are widespread and can be easily collected. However, for programs with non-standard input formats or even self-customized formats, this strategy becomes unaffordable. An alternative solution is run \textit{generators} that can produce test cases automatically. This only works when an applicable generator already exists. Otherwise, new generators must be built from scratch, which, as we will elaborate in~\S\ref{subsec:bcg:seedgen}, is often infeasible or impractical.

The recent development of AI,  especially Large Language Models (LLMs) such as the Generative Pre-trained Transformer (GPT)~\cite{floridi2020gpt}, offers a new opportunity for generic, effortless seed preparation. Many LLMs (GitHub Copilot~\cite{copilot}, Amazon CodeWhisperer~\cite{codewhisperer}, OpenAI's GPT series~\cite{openaimodels}, Anthropic's Claude series~\cite{Anthropicmodels}, etc.), after pre-training on massive code, comments, and documentation, are excelling at code-centric tasks (code understanding, code summarization, code completion, code translation, etc.)~\cite{lu2021codexglue,nam2024using,chen2021evaluating,wang2023codet5+}. A straightforward application of the technology would be to run an LLM to analyze the target program for fuzzing and generate the test cases it expects.

Indeed, recent research has explored LLMs for seed generation in fuzzing~\cite{ackerman2023large,xia2024fuzz4all,lohiya2024poster,tamminga2023utilizing,liu2024llm,rutherford2003case,yang2023kernelgpt}. The proposed methods have attempted using various inputs (target program code, example test cases, functionality specifications, documentation, etc.) and prompts to request test cases from the LLMs. They have demonstrated effectiveness within different domains (parsers~\cite{ackerman2023large}, compilers~\cite{xia2024fuzz4all}, the Linux kernel~\cite{yang2023kernelgpt}, and other generic software~\cite{liu2024llm}). However, they may have significantly undermined the potential of LLMs for seed generation due to failure to address several fundamental challenges. \textbf{(C1)} The LLM in use, even in their most recent versions, may not support many input formats. For instance, GPT-4o~\cite{gpt4o}, OpenAI's new flagship model, refuses to generate binary representations because it is constrained to text formats. \textbf{(C2)} LLMs are restricted by their \textit{context window}---the amount of token the model can handle from both its input and response~\cite{contextwindow}. Arbitrarily dumping information (e.g., the entire code base of the target program) to the LLMs can overflow the context window and fail the generation.  \textbf{(C3)} LLMs are known to present unpredictable behaviors, which can impede the generation of test cases. \textbf{(C4)} The LLMs can lack a basic understanding of the progress. Thus, it may overlook an incomplete task or endlessly repeat an accomplished task.

In this paper, we present a system, \sys, as an attempt toward generic, effective LLM-based seed generation. \sys incorporates four ideas to address challenges \textbf{C1 - C4}. \ding{182} Inspired by a recent OSS-Fuzz extension~\cite{ossfuzzgen}, \sys instructs the LLM to construct a generator that can produce test cases instead of asking it to directly output test cases. The generator can be represented as pure text, which, when executed, can generate seeds in any format (\textbf{overcoming C1}). \ding{183} \sys devises coverage-guided evolution by incorporating a feedback-based loop to guide the LLM to improve the generator toward broader and deeper code coverage gradually. This enables guided progress, avoiding blind explorations (\textbf{overcoming C4}). \ding{184} \sys only provides the LLM with the context necessary to improve the generator rather than dumping everything (e.g., all the code of the target program), avoiding overflowing the context window (\textbf{overcoming C2}). \ding{185} \sys introduces state-driven realignment. Once observing LLM behaviors deviating from the expected state, \sys attempts to re-align the LLM in the right direction via behavior-amending instructions (\textbf{overcoming C3}).

To understand the utility of \sys, we have performed an intensive set of evaluations. \ding{192} We apply \sys to generate test cases for all the functional C and C++ targets included in the OSS-Fuzz project~\cite{ossfuzzrepo} (\ossfuzzprog programs and \ossfuzzharness harnesses). It shows that \sys can generate seeds with a quality close to the human-created ones. Further, \sys significantly outperforms the existing LLM-based solution. \ding{193} We use \sys to prepare seeds for running  \afl~\cite{fioraldi2023dissecting}, AFL++~\cite{fioraldi2020afl++}, \honggfuzz~\cite{honggfuzz} on \magma~\cite{hazimeh2020magma}, a ground-truth fuzzing evaluation suite based on real programs with real bugs. The results demonstrate that \sys is compared to human-created seeds in facilitating fuzzing to find bugs. Likewise, \sys beats the existing LLM-based solution to a remarkable extent. \ding{194} We diversify the LLM used by \sys to span \gptfo~\cite{gpt4o}, \gptturbo~\cite{gpt35turbo}, and \claudesonnet~\cite{claude35sonnet}. \sys presents effectiveness with all three models, showing its generality. \ding{195} We measure the cost of \sys for seed generation. \sys can generate high-quality seeds for a fuzzing harness with an average cost of less than \$0.5, showing its affordability. We have also applied \sys in DARPA and ARPA-H’s Artificial Intelligence Cyber Challenge (AIxCC)~\cite{aixcc}, a world-level, cutting-edge competition on developing AI-powered solutions for securing critical software infrastructures. \sys aided us in becoming one of the leading teams.

In summary, our main contributions are as follows.

\begin{itemize}[leftmargin=2em]
\setlength{\itemsep}{4pt}
\item We introduce \sys, a system to offer generic, effective LLM-based seed generation.
\item We incorporate a group of new ideas into \sys for addressing the major, prevalent challenges encountered by LLM-based seed generation.
\item We implement \sys and intensively evaluate \sys on standard benchmarks. The results show that \sys offers generic, effective, and economical seed generation.
\end{itemize}

%% file: 02_background.tex
\section{Background\label{sec:bcg}}

\subsection{Greybox Fuzzing\label{subsec:bcg:grexboxfuzzing}} 

Greybox fuzzing~\cite{li2018fuzzing,manes2018art,godefroid2020fuzzing} emerged in the mid-2000s as a balance between blackbox~\cite{godefroid2007random} and whitebox fuzzing~\cite{godefroid2012sage}. The concept was popularized by tools such as American Fuzzy Lop (AFL)~\cite{fioraldi2023dissecting}. The key idea of greybox fuzzing is to use lightweight instrumentation to obtain runtime feedback (e.g., code coverage) during fuzzing and, guided by the feedback, pick and mutate the existing test cases to derive new ones. To close the loop, new test cases offering new contributions (e.g., covering new code) are preserved for future mutations.

A greybox fuzzing campaign usually starts with an initial corpus of test cases called \textit{seeds}. The quality of seeds significantly influences the fuzzing performance. For instance, given seeds covering a large code region, the chance of fuzzing to discover more bugs will be much higher, and the time needed will be much shorter. Hence, preparing a set of high-quality seeds has become a de facto standard before launching greybox fuzzing.

\subsection{Seed Generation\label{subsec:bcg:seedgen}}

To prepare high-quality seeds, a common practice is to manually understand the target program and craft desired test cases. However, this is not ideal for the highly automated process of fuzzing. An alternative idea is to run \textit{generators} that can produce test cases automatically. Yet, it faces the challenge of constructing a generator when none is available.

To date, there are two main methods for constructing generators. \ding{182} Given the format specifications of input needed by the target program, people manually develop generators complying with the specifications to assemble test cases~\cite{xu2020freedom,domato,yang2011finding}. Those generators have mostly targeted standard formats like XML, HTML, and MathML, as their specifications are well documented. \ding{183} The other approach is based on machine learning techniques. After gathering a significant volume of inputs accepted by the target program, the method trains a model (probabilistic models~\cite{wang2017skyfire}, recurrent neural networks~\cite{godefroid2017learn}, generative adversarial networks~\cite{lyu2018smartseed}, etc.) to learn the input structures, which is then run to generate new input variants.

Evidently, the two methods above are not preferable. They still \textit{mandate manual efforts} to engineer the generator or collect the training data. More fundamentally, \textit{they lack generality}. When the target program requires input following a non-standard or brand-new format, both methods become impractical. On the one side, format documentation is absent. Reversing the program to infer the format and building generators accordingly is unaffordable. On the other hand, satisfactory inputs in the wild become rare. There are limited sources to build an effective training dataset.

\subsection{AI for Code\label{subsec:bcg:aicodegen}}

The recent development of large language models (LLMs)~\cite{chang2024survey} has revolutionized the capability of AI in understanding and generating structured and unstructured text. Many models (e.g., GitHub Copilot~\cite{copilot}, OpenAI Codex~\cite{codex}, Amazon CodeWhisperer~\cite{codewhisperer}) are specifically designed and adapted for working with code. They present unprecedented performance in various code-centric tasks (code summarization, code completion, code translation, etc.)~\cite{lu2021codexglue,nam2024using,chen2021evaluating,wang2023codet5+}. Even generic models like OpenAI's GPT family and Anthropic's Claude family, after pre-training on massive code, comments, and documentation, are excelling at those tasks~\cite{lu2021codexglue}.

The strong code capability of LLMs has spurred their use in seed generation for fuzzing~\cite{ackerman2023large,xia2024fuzz4all,lohiya2024poster,tamminga2023utilizing,liu2024llm,rutherford2003case,yang2023kernelgpt}. The existing methods in this line provide various types of input (target program code, example test cases, functionality specifications, documentation, etc.) together with customized prompts to LLMs, asking the LLMs to output test cases. They have demonstrated effectiveness on programs in various domains, such as parsers~\cite{ackerman2023large}, compilers~\cite{xia2024fuzz4all}, the Linux kernel~\cite{yang2023kernelgpt}, and other generic software~\cite{liu2024llm}. Compared to the aforementioned approaches for seed generation, LLM-based methods require less human effort while offer better generality. Yet, the existing methods have overlooked several challenges we will discuss shortly in~\S\ref{subsec:techchallenge}. As a result, they have insufficiently utilized the potential of LLMs.

%% file: 03_challenges.tex
\section{Problem and Challenges\label{sec:challenge}}

\begin{figure}[t]
    \centering
    \begin{tcolorbox}[colback=white, colframe=blue!60!black, title={Prompt: \textit{Give me the actual binary representation for ``ELF''}}]
     \textbf{\gptfo:} \\
    ``\textit{I understand you're asking for an actual binary sequence (not just a text-based representation of it). However, since I can only communicate using text, I cannot provide literal binary byte streams like a file would contain, as my output is constrained to text formats.}''
    \end{tcolorbox}
    \vspace{-1em}
    \caption{Illustration of challenges incurred by input formats.}
    \label{fig:formatchallenge}
\end{figure}

\subsection{Problem Statement\label{subsec:problem}}

In this paper, we focus on exploiting LLMs for seed generation in greybox fuzzing. To better define the problem, we clarify our assumptions below:

\begin{itemize}[leftmargin=2em]
\setlength{\itemsep}{4pt}
\item \textbf{Input:} We assume access to the source code of the target program. We also assume the entry point of the fuzzing target is specified (e.g., the entry function needed by \libfuzzer~\cite{libfuzzer}). The two assumptions are minimal for functional fuzzing. Compared to the previous methods, we enforce no extra requirements (e.g., the availability of functionality specifications and documentation), thus representing a more generic scenario.

\item \textbf{LLM:} We only require black-box access to the LLM. That is, the LLM can take our prompts as inputs and send us responses. The responses can be purely text-based, and support for other formats (e.g., image, audio, video) is not demanded. 

\item \textbf{Goal:} Previous research on LLM-based seed generation usually follows a vague objective like generating more diverse test cases. We consider a more specific goal. \textit{We aim to generate test cases maximizing the code coverage in the target program}. The rationale is that modern greybox fuzzing tools prevalently take code coverage as their top priority. Hence, fulfilling our goal will better assist with the fuzzing process.
\end{itemize}

\subsection{Technical Challenges\label{subsec:techchallenge}}

LLM-based seed generation appears to be intuitive. We may engineer a prompt, enclosing the target program code and the entry point information, to ask the LLM to produce diversified test cases. Indeed, previous research~\cite{lohiya2024poster} has applied this idea. However, it tremendously undermines the potential of LLMs for seed generation, due to its failure to address several fundamental challenges.

\vspace{0.5em}
\noindent\textbf{Heterogeneous Input Formats (C1):} The target programs of fuzzing can require a variety of input formats. For instance, the OSS-Fuzz project~\cite{serebryany2017oss} has included 1260 open-source programs, spanning 130 input formats (text, image, video, audio, binary, etc.). \textit{The LLM in use, even in their most recent versions, may not support many of those formats}. For instance, ~\autoref{fig:formatchallenge} illustrates that GPT-4o~\cite{gpt4o}, OpenAI's new flagship model, refuses to generate binary representations because it is constrained to text formats. In short, asking the LLM to directly generate desired test cases can fail.

\vspace{0.5em}
\noindent\textbf{Limited Context Window (C2):} Restricted by computational resources and model architecture designs, LLM typically enforces a \textit{context window}---the amount of token the model can handle from both its input and response~\cite{contextwindow}. For instance, GPT-4o has a context window of 128k tokens, while GPT-3.5-Turbo only has 16k. \textit{Larger programs, such as server programs and OS kernels, can easily have a code size exceeding the context window, failing to be consumed by the LLMs}. Even given a sufficiently large context window, it is unwise to feed all code to the LLM, as a longer context can degrade the LLM's reasoning capability~\cite{contextwindowreasoning}.

\vspace{0.5em}
\noindent\textbf{Unpredictable LLM Behaviors (C3):} LLM is known to present unpredictable behaviors. \textit{Such behaviors can impede the generation of test cases}. For instance, the LLM can run into hallucinations~\cite{yao2023llm}, generating test cases that seem plausible but unacceptable to the target program. It can also carry bias~\cite{huang2023bias}, preferring test cases with identical properties and compromising the diversity. 

\vspace{0.5em}
\noindent\textbf{Blind Space Exploration (C4):} In essence, the LLM needs to explore the code space of the target program and generate test cases covering different code blocks. Yet, by merely looking at the code, \textit{the LLM lacks a basic understanding of the progress}. It may mistakenly believe a new block has been covered and skip it, or it may endlessly work on a block even if it has been covered, leading to limited yields or a waste of resources.

%% file: 04_design.tex
\section{Design and Implementation\label{sec:design}}

\begin{figure*}[!t]
  \centering
  \scriptsize

    \includegraphics[width=\textwidth]{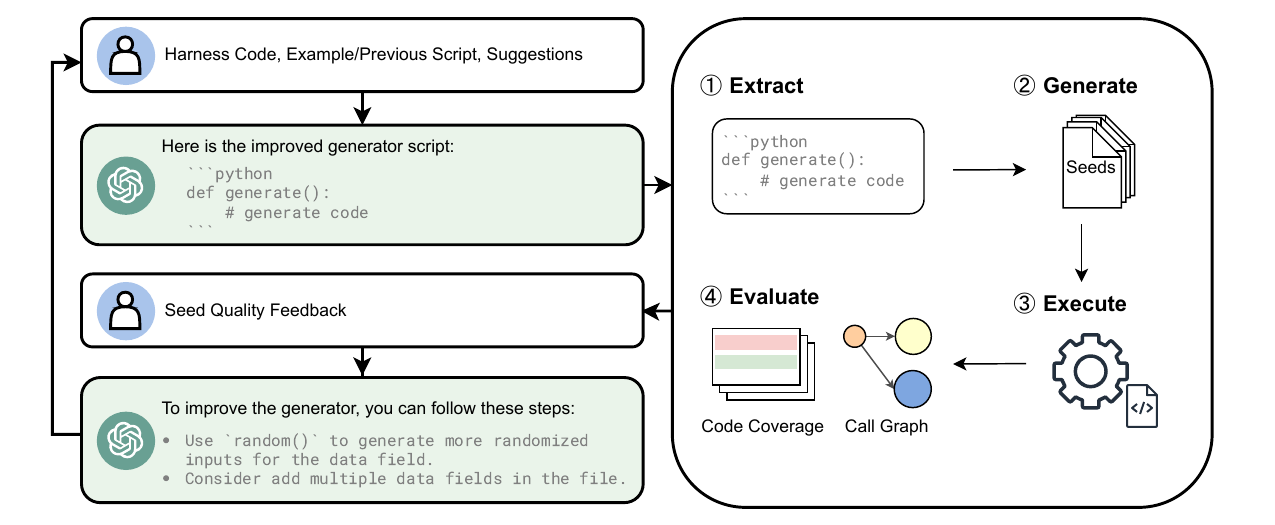}
    \vspace{-1.15em}
    \caption{Workflow of \sys.\label{fig:framework}}
  \vspace{-0.75em}
  \label{fig:workflow}
\end{figure*}

\subsection{Overview}
In this section, we present our system, \sys, as the first attempt to address challenges \textbf{C1 - C4}. The workflow of \sys is shown in~\autoref{fig:workflow}. Instead of asking the LLM to directly output test cases, \sys instructs the model to construct a generator that can produce test cases, following inspiration from a recent OSS-Fuzz extension~\cite{ossfuzzgen}. The generator, consisting of only code, can be represented as pure text. Yet, via execution, it can generate test cases of any format. This overcomes LLM's restriction on response format (\textbf{addressing C1}). 

At its core, \sys incorporates a feedback-based loop to guide the LLM to gradually improve the generator toward producing test cases with broader and deeper code coverage. This design enables the LLM to explore the code space of the target program in a guided manner, mitigating blind explorations (\textbf{addressing C4}). To avoid overflowing the context window, we only provide the LLM with the context necessary to improve the generator rather than dumping everything (e.g., all the code of the target program) to it (\textbf{addressing C2}).

While guiding the LLM to improve the generator, \sys tracks its behaviors. Once observing behaviors deviating from the expected state, \sys attempts to re-align the LLM in the right direction via behavior-amending instructions (\textbf{addressing C3}). In the following, we elaborate on the technical details of \sys.

\subsection{Initial Generator}

Given a fuzzing target, \sys starts with producing a basic while functional generator. It first identifies the entry function of fuzzing and extracts its source code. For instance, given a fuzzing target prepared as a \libfuzzer harness (one of the standard and most popular formats today)~\cite{libfuzzer}, \sys considers \code{LLVMFuzzerTestOneInput()} as the entry function. Using an elaborately crafted prompt which includes the entry function as a piece of context, \sys requests the LLM to output a test case generator in Python. 

The prompts \sys adopted are presented in Appendix~\ref{subsec:prompt}. While prompt engineering is not a focus of our research, those prompts are the results of numerous adaptations and optimizations in AIxCC~\cite{aixcc}, a world-leading competition on developing AI-powered solutions for securing critical software infrastructures. They have presented both generality and robustness in various real-world applications.

LLMs occasionally fail to produce a parsable and executable generator. We consider this an unpredictable behavior, and we will explain how we address it shortly in~\S\ref{subsec:realignment}.

\begin{figure*}[!t]
  \centering
  \scriptsize
    \includegraphics[width=1\textwidth]{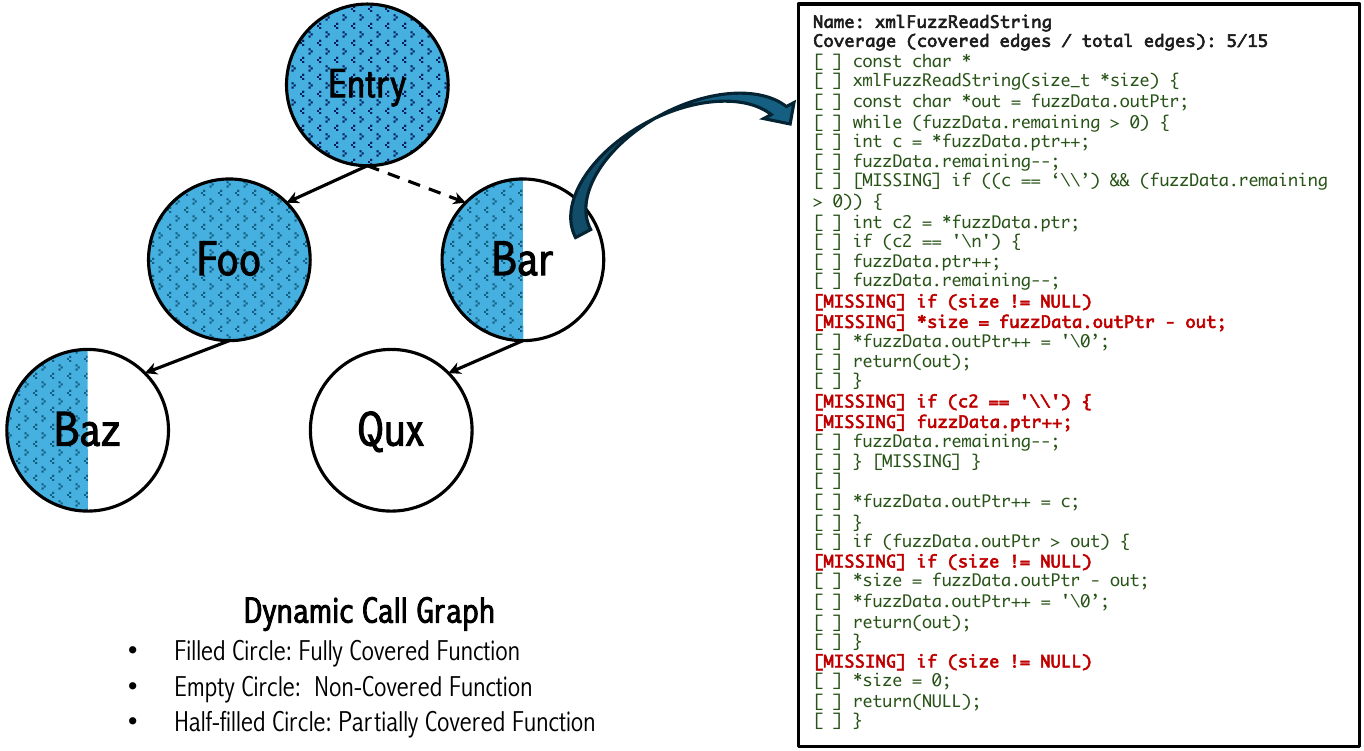}
    \vspace{-1em}
    \caption{An illustration of code coverage on dynamic call graph.\label{fig:cg}}
  \vspace{-0.75em}
  \label{fig:codecov}
\end{figure*}

\subsection{Coverage-Guided Evolution} Even when the initial generator can run without issues, it oftentimes only generates test cases covering a limited region of code. We design a coverage-guided strategy to evolve the generator. Instead of aiming for a single, ultimate generator to reach maximal code coverage, we guide the LLM to iteratively create new variants to reach the non-covered code piece by piece.

\vspace{0.5em}
\noindent\textbf{Code Coverage Collection:} Once a new working generator $\mathcal{G}$ is produced by the LLM, we run it to generate $N$ (1,000 by default\footnote{We observe that a single generator created by LLMs usually present limited diversity. Running it to generate 1,000 test cases can usually reach its maximal capability. 1,000 is also the default used by OSS-Fuzz's AI-based seed generator~\cite{ossfuzzgen}.}) test cases. The code coverage of the test cases, quantified by branch coverage, is measured and merged with that of all previous generators. The total code coverage is represented on a \textit{dynamic call graph} (i.e., the call graph composed of functions visited by all test cases from the existing generators and their children functions). ~\autoref{fig:codecov} illustrates this representation of code coverage.

\vspace{0.5em}
\noindent\textbf{Prompt Assemble:}\label{subsec:design:prompt_assemble}
To evolve our generator $\mathcal{G}$, we aim to create a new variant that can reach the uncovered code segments. We use the most recent version of $\mathcal{G}$ along with the updated call graph and code coverage information to assemble a prompt. This prompt is then fed to the LLM as a feedback on its progress. However, a challenge arises: the call graph can be extensive, potentially resulting in a prompt that exceeds the LLM's context window limitations.

To address the challenge of context window limitations, we employ two key strategies for optimizing context usage. First, we focus exclusively on partially covered functions in the call graph (illustrated in the right side of~\autoref{fig:codecov}). We exclude fully covered and non-covered functions to reduce context size and maintain relevance. This exclusion is justified because (i) non-covered functions, being descendants of partially covered ones, do not contribute to covering missed branches in their predecessors, and (ii) $\mathcal{G}$ already incorporates knowledge about fully covered functions, which is provided to the LLM.

If the context still exceeds the LLM's capacity after this initial pruning, we implement an iterative pruning approach based on the dynamic call graph. Starting from the deepest level of the call graph, we progressively remove functions and move upwards until the token count falls within the specified limit. This method ensures efficient utilization of the available context while preserving the most critical functions in the program's execution flow.

\vspace{0.5em}
\noindent\textbf{New Generator Creation:}
After optimizing the context using the strategies described above, we feed the resulting prompt to the LLM to obtain suggestions for improving $\mathcal{G}$. This step is crucial as it guides the LLM to focus specifically on enhancing the generator's capabilities. An example of such a suggestion is illustrated in~\autoref{fig:workflow}. By explicitly requesting suggestions before asking for a new generator, we create a more focused chain-of-thought~\cite{wei2022chain}, ensuring the LLM concentrates on generator improvement rather than being distracted by other tasks.

These suggestions serve as valuable input for the subsequent iteration. We incorporate them into a new prompt, along with $\mathcal{G}$ and the partially covered functions identified earlier. This comprehensive prompt is then used to request an improved generator from the LLM.

\subsection{State-Driven Realignment\label{subsec:realignment}}

LLMs can occasionally produce outputs that do not meet our requirements or exhibit unexpected behaviors. For instance, they may generate scripts that cannot be executed or fail to produce a script altogether. To address these issues, we employ a state-driven framework to regulate the system's behavior and ensure more consistent and reliable outputs.

\begin{figure*}[ht]
  \centering
  \scriptsize
    \includegraphics[width=\textwidth]{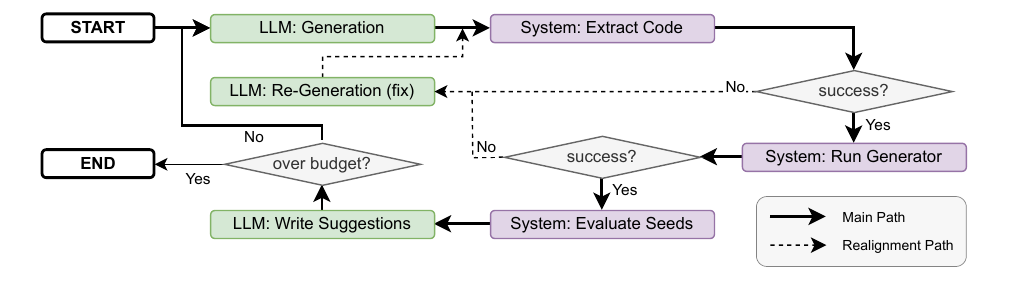}
    \vspace{-4em}
    \caption{State Machine of \sys.}
  \vspace{-0.75em}
  \label{fig:state_machine}
  \vspace{-1em}
\end{figure*}

As is shown in~\autoref{fig:state_machine}, \sys employs a state machine to manage the progress of evolution. When the system detects unpredictable behaviors from the LLM, it initiates a realignment process with behavior-amending instructions. These instructions reverts the system to a previous state and prompting the LLM to rectify the error. For instance, if a generator script fails to execute, the system captures the standard error output and stack trace. This information is then fed back to the LLM with a request to diagnose and resolve the issue.

\subsection{Implementation}

\sys consists of two main components: an LLM agent and a runtime daemon. Our implementation comprises 5,162 lines of code in total, distributed across three programming languages: 2,096 lines of Go, 1,643 lines of Python, and 1,423 lines of Rust.

\vspace{0.5em}
\noindent\textbf{LLM Agent:} The LLM agent is implemented in Python using the LangGraph framework. It operates as a state machine, with each state represented as a LangGraph node and transitions between states as edges in the graph. The state machine encompasses the processes of initial generator creation, coverage-guided evolution, and state-driven realignment.

\vspace{0.5em}
\noindent\textbf{Runtime Daemon:} The runtime daemon is written in Go and statically linked. It is designed for injection into isolated environments such as Docker containers used in OSS-Fuzz projects. The daemon's functions include compiling the target code, executing the target with generated seeds, collecting code coverage and function calling information, and generating the dynamic call graph.

\vspace{0.5em}
\noindent\textbf{Code Coverage and Call Graph Generation:} Code coverage and function calling information are collected using multiple tools. A custom LLVM pass is used to instrument the program for dynamic function call data collection. Concurrently, LLVM's Coverage Sanitizer gathers code coverage data. To align this coverage information with the source code, tools such as \code{nm} and \code{llvm-symbolizer} are utilized to locate and extract relevant code snippets from source files.

\vspace{0.5em}
\noindent\textbf{Integration and Workflow:} The LLM agent controls the overall process, determining generator improvements and realignment strategies. The runtime daemon executes these decisions in the target environment and provides coverage data and execution results back to the agent. Communication between the LLM agent and the runtime daemon is implemented using gRPC.

%% file: 05_evaluation.tex
\section{Evaluation\label{sec:eval}}
To assess the utility of \sys, we perform a series of evaluations focusing on five questions:

\vspace{0.5em}
\noindent\ding{192} \textit{Can \sys generate high-quality seeds?}

\vspace{0.5em}
\noindent\ding{193} \textit{Can seeds generated by \sys facilitate fuzzing?}

\vspace{0.5em}
\noindent\ding{194} \textit{Can \sys outperform the existing LLM-based solutions?}

\vspace{0.5em}
\noindent\ding{195} \textit{What is the impact of the LLM used by \sys?}

\vspace{0.5em}
\noindent\ding{196} \textit{What is \sys's cost to generate seeds for a fuzzing target?}

\subsection{Experimental Setup\label{subsec:eval:setup}}

\vspace{0.5em}
\noindent{\bf Benchmarks:} To support our evaluation, we adopt two benchmarks. The first benchmark includes open-source programs collected from the OSS-Fuzz project~\cite{ossfuzzrepo}. We consider all the C and C++ programs in OSS-Fuzz. After excluding those without a seed corpus or unable to work with OSS-Fuzz's coverage utility\footnote{\url{https://github.com/google/oss-fuzz/blob/master/infra/base-images/base-runner/coverage}}, we end up with \ossfuzzprog programs. Each program configures one or more fuzzing harnesses following the format specified by~\libfuzzer~\cite{libfuzzer}. In total, we include \ossfuzzharness harnesses. The goal of this benchmark is to asses whether \sys can generate high-quality test cases for a wide spectrum of programs.

The second benchmark is \magma~\cite{hazimeh2020magma}, a ground-truth fuzzing evaluation suite based on real programs with real bugs. It includes 9 widely used open-source projects and 16 \libfuzzer-style harnesses. We include this benchmark to measure how much \sys can facilitate fuzzing in discovering real-world bugs.

\vspace{0.5em}
\noindent{\bf Baselines:} We consider two baselines of seed preparation. The first baseline is the seed corpus shipped with both benchmarks\footnote{An example of default corpus shipped with OSS-Fuzz: \url{https://github.com/DavidKorczynski/binary-samples}; \\An example of default corpus shipped with \magma: \url{https://github.com/HexHive/magma/tree/v1.2/targets/libxml2/corpus/libxml2_xml_read_memory_fuzzer}.}. The seeds were manually collected and maintained by the benchmark developers. They are also plentiful in number, ranging from dozens to thousands for each harness.  This baseline can well represent the high-quality people manually prepare for greybox fuzzing. The second baseline is an AI-based seed generation solution Google recently extended for OSS-Fuzz~\cite{ossfuzzgen}. It uses the basic idea of providing an LLM with the code of a fuzzing harness and asking the LLM to output test case generators in Python. This baseline represents a weaker version of LLM-based methods. For simplicity, we use \ossai to refer to this baseline. For the \magma benchmark, we additionally include an empty seed corpus as a third baseline, simulating a scenario where the fuzzing users do not prepare seeds.

\vspace{0.5em}
\noindent\textbf{Experiment Configurations:} We conduct all experiments on CloudLab~\cite{duplyakin2019design}, with each machine equipped with Intel Skylake processors (20 physical cores @ 2.20GHz) and 192GB of RAM. For evaluations on the OSS-Fuzz benchmark, we run \sys and \ossai for 30 minutes on each fuzzing harness. We further limit the LLM API fees to \$0.5 per hardness to avoid cost explosion\footnote{We cannot restrict the API fees of \ossai as it uses Google's proxy, which offers no interfaces to retrieve the costs.}. On a specific harness, we run each generator to produce 1,000 test cases, and we set it to time out after 30 seconds to mitigate non-terminating generators.

For evaluations on the \magma benchmark, we respectively run \afl~\cite{fioraldi2023dissecting}, AFL++~\cite{fioraldi2020afl++}, \honggfuzz~\cite{honggfuzz} on each hardness with seeds from \sys and the three baselines (i.e., seeds generated by \ossai, the default seed corpus, and the empty seed corpus). Each harness is run for 24 hours with 3 instances affiliated to separate physical CPU cores. Each run is repeated 5 times to neutralize randomness. 

To understand the impacts of different LLMs, we repeat all the OSS-Fuzz experiments with \sys and \ossai, separately using \gptfo~\cite{gpt4o}, \gptturbo~\cite{gpt35turbo}, and \claudesonnet~\cite{claude35sonnet} as the LLM. We also employ a two-tier isolation strategy to ensure standardized and controlled testing conditions. First, for each fuzzing target, we run \sys in an isolated environment. This setup allows for independent execution of the LLM agent, enhancing security by isolating untrusted code generation. Second, each fuzzing target operates within its own Docker container, ensuring that each target has access to its required compilation environment and runtime libraries without interference from other targets.

\vspace{0.5em}
\noindent\textbf{Evaluation Metrics:} For the OSS-Fuzz benchmark, we consider the code coverage, measured by branch coverage, of the seeds as the performance metric. For the \magma benchmark, we reuse the metrics recommended by the maintainers~\cite{hazimeh2020magma}, including the time to reach a bug (TR) and the time to trigger a bug (TT).

\begin{figure*}[!t]
 \centering
 \scriptsize
   \includegraphics[width=\textwidth]{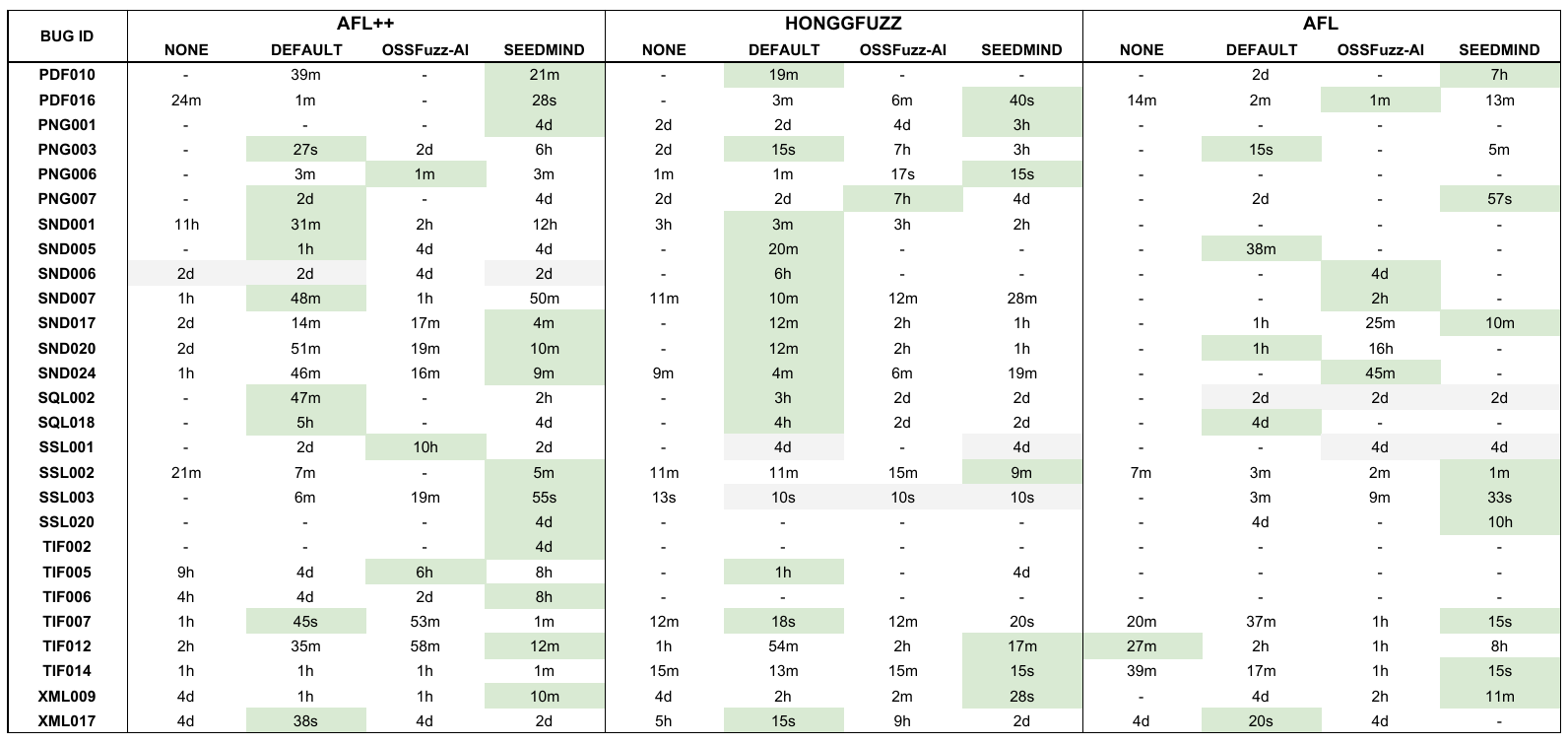}
   \vspace{-2em}
   \caption{Results of bug-finding evaluation with MAGMA. \textbf{NONE} means no seeds are used, and \textbf{DEFAULT} represents the default seed corpus shipped with the fuzzing target. The numbers stand for the average time-to-trigger of the corresponding bug. Values highlighted with \textcolor{cadmiumgreen}{green} indicate the shortest time-to-trigger among the four solutions.}
 \vspace{-1em}
 \label{fig:magma}
\end{figure*}

\subsection{Quality of Test Cases}
To understand the quality of the test cases generated by \sys, we inspect their code coverage in the OSS-Fuzz programs. The detailed results are presented in Appendix~\S\ref{subsec:eval:codecov}. In summary, \sys can generate test cases to achieve satisfactory code coverage.

\vspace{0.5em}
\noindent\textbf{Comparing with Default Corpus:} Among the \ossfuzzharness harnesses, \sys achieves greater code coverage than the default seed corpus on 48 harnesses when using \gptturbo, 253 harnesses with \gptfo, and 268 harnesses with \claudesonnet. For these specific harnesses, \sys's code coverage is 43.0\%, 44.1\%, and 39.7\% higher than the default seed corpus, respectively. Taking all the harnesses into account, \sys achieves 72.0\%, 89.3\%, 87.7\% of the code coverage reached by the default seed corpus when using \gptturbo, \gptfo, and \claudesonnet as the LLM. The results show that, \textit{while not fully comparable, the seeds generated by \sys present quality close to that of the human-created corpus}. In many cases, \sys can even offer advantages. 

\vspace{0.5em}
\noindent\textbf{Comparing with \ossai:} Both as LLM-based solutions, \textit{\sys significantly outperforms \ossai}. Regardless of which LLM is used, \sys achieves higher code coverage for a substantially larger number of harnesses than \ossai. With \gptturbo, \sys and \ossai can both run on 159 harnesses and \sys reaches higher code coverage on 142 of them. Switching to \gptfo and \claudesonnet, \sys outperforms \ossai on 537 out of 636 harnesses and 588 out of 674 harnesses, respectively. If we only look at those harnesses, \sys presents code coverage 29.0\%, 23.6\%, 23.3\% higher than \ossai. In the few instances where \ossai covers more code, the difference is mostly marginal and negligible. Thus, with all harnesses counted, \sys still shows code coverage 27.5\%, 23.6\%, 23.3\% higher than \ossai across the three LLM configurations. 

\subsection{Benefits to Fuzzing}

For assessing how much the seeds generated by \sys benefit fuzzing, we compare the results of \sys and the three baselines discussed in~\S\ref{subsec:eval:setup} on \magma. In this evaluation, we fix \sys and \ossai to use \claudesonnet because, as we will show shortly, \claudesonnet enables the best performance. Overall, \sys can facilitate the bug finding of all three fuzzing tools. For simplicity of presentation in the following, if a solution finds a bug using a time shorter than all other solutions, we call the bug a \textit{fastest bug} found by that solution.

\vspace{0.5em}
\noindent\textbf{Comparing with Default Corpus:} \sys appears comparable to the default seed corpus when applied for bug finding. Both present varying but close performances on different fuzzing tools. With \aflpp, \sys enables the discovery of 27 bugs, and the default corpus enables 24. In addition, \sys finds 13 fastest bugs, while the default corpus finds 9. With \honggfuzz, the results are invested. The default corpus enables 24 bugs, with 14 fastest bugs. \sys only enables 21 bugs, with 7 fastest bugs. Their performance with \afl is more consistent. The default corpus enables more bugs (17 \textit{v.s.} 14), but \sys finds more fastest bugs (9 \textit{v.s.} 5).

\vspace{0.5em}
\noindent\textbf{Comparing with \ossai:} \sys clearly beats \ossai. With \aflpp, \sys enables the discovery of 9 more bugs (27 \textit{v.s.} 18). \sys also finds bugs faster (13 fastest bugs \textit{v.s.} 9 fastest bugs). The results with \honggfuzz are similar. The gap with \afl is smaller. They find the same amount of bugs, but \sys has a much faster pace (9 fastest bugs \textit{v.s.} 4 fastest bugs).

\vspace{0.5em}
\noindent\textbf{Comparing with Empty Corpus:} \sys thoroughly defeats the empty corpus by consistently finding more bugs (27 \textit{v.s.} 15 with \aflpp, 21 \textit{v.s.} 14 with \honggfuzz, and 14 \textit{v.s.} 6 with \afl) at a fast pace (29 fastest bugs in total \textit{v.s.} 1fastest bug in total). These show that \sys is a promising alternative when no default corpus is available.

\subsection{Generality to LLM}

To assess the generality of \sys across different language models, we conducted experiments using three tiers of LLMs: \gptturbo, \gptfo, and \claudesonnet. Our results summarized in~\autoref{tab:compmodel} demonstrate that \sys presents effectiveness with all three models, showcasing its adaptability to various LLM architectures and capabilities.

\input{tables/models-comp}

While \sys proves functional across all tested LLMs, we observed notable performance discrepancies. \claudesonnet results in superior performance, leading in code coverage for 385 harnesses and achieving an average code coverage of 18.03\%. This is followed by \gptfo, leading in 272 harnesses with an average coverage of 17.12\%. In contrast, \gptturbo only achieves an average coverage of 15.12\%, leading in 17 harnesses.

These results indicate that while \sys is effective with all tested LLMs, its performance can be enhanced by using more advanced models.
We observed a positive correlation between the context window size and performance. \claudesonnet, with the largest context window of 200,000 tokens, outperforms its counterparts, while \gptturbo, with the smallest window of 16,385 tokens, performs less. This is potentially because our pruning strategy described in~\S\ref{subsec:design:prompt_assemble} is applied more aggressively when the context window is smaller, as in \gptturbo.

Other factors may also contribute to these performance disparities. For example, the superior results of \claudesonnet and \gptfo could be attributed to their lager model sizes and more diverse training datasets, enabling them to generate more effective and varied test cases.

\subsection{Cost Analysis}

We conducted a cost analysis to evaluate the economic feasibility of \sys for practical use. As explained before, we enforce a soft upper bound of \$0.5 per harness to manage costs. This approach involves checking the accumulated cost after each iteration of seed generation. If the cost has not exceeded \$0.5, the system continues to the next iteration. This method allows for slight budget overruns, ensuring that the last valuable iteration is not cut short.

\input{tables/costs}

Table~\ref{tab:costs} presents the average cost per fuzzing harness for each LLM, along with the number of harnesses that remained within our \$0.5 bound. \claudesonnet strikes a balance between cost and performance, with an average cost of \$0.48 per harness. It managed to stay within the \$0.5 budget for 206 harnesses, the highest among all models, indicating its consistent performance across a wide range of scenarios. \gptfo, while slightly exceeding our soft upper bound with an average cost of \$0.69, remains acceptable given its strong performance.

It's noteworthy that for a significant number of harnesses, all models remained under the \$0.5 threshold (49 for \gptfo, 159 for \gptturbo, and 206 for \claudesonnet) within a strict 30-minute time limit. This suggests that \sys can be deployed cost-effectively for many fuzzing tasks, with the flexibility to allocate more resources to complex harnesses when necessary.

%% file: tables/models-comp.tex
\begin{table*}[!t]
  \caption{Comparison of models}
  \vspace{-0.75em}
  \label{tab:compmodel}
  \begin{tabular}{lccc}
\hline
\textbf{Model} & \textbf{Context Window} & \textbf{\# of Highest Coverage} & \textbf{Average Coverage \%} \\
\hline
\gptturbo & 16,385 tokens &  17 & 15.12 \\
\hline
\gptfo & 128,000 tokens & 272 & 17.12 \\
\hline
\claudesonnet & 200,000 tokens & 385 & 18.03 \\
\hline
\end{tabular}
\end{table*}

%% file: tables/costs.tex
\begin{table*}[!t]
  \caption{Average cost of running \sys to generate seeds for a fuzzing harness.}
  \label{tab:costs}
  \vspace{-0.75em}
  \begin{tabular}{lcc}
\hline
\textbf{Model} & \textbf{Average Cost \$} & \textbf{\# of harnesses (<0.5\$)} \\
\hline
\gptfo & 0.69 & 49 \\
\hline
\gptturbo & 0.10 & 159 \\
\hline
\claudesonnet & 0.48 & 206 \\
\hline
\end{tabular}
\vspace{-1em}
\end{table*}

%% file: 06_relatedwork.tex
\section{Related Work\label{sec:related}}

\subsection{Seed Generation for Fuzzing}

Generation-based fuzzing can produce highly structured inputs for real-world applications. Various approaches for structured test case generation have evolved over time:

\vspace{0.25em}
\noindent\textbf{Manually summarizing grammar rules.} Generation-based fuzzers require well-written grammar rules prior to generating test cases. Examples of such fuzzers, designed for producing syntax-correct HTML files, include DOMATO~\cite{sysmeon}, FREEDOM~\cite{xu2020freedom}, and DOMFUZZ~\cite{mozillafuzzing}. DOMFUZZ also employs a grammar-based splicing technique, which inspires our hierarchy object exchanging method. For fuzzing JavaScript codes, techniques like ~\cite{park2020fuzzing}, ~\cite{mozillafuzzing1}, and \cite{holler2012fuzzing} use random generation or combination of code based on provided syntax rules. Favocado~\cite{Dinh2021FavocadoFT} generates syntactically correct binding code for fuzzing JavaScript engines using semantic information.

\vspace{0.25em}
\noindent\textbf{Grammar generation with machine learning.} Learn-\&Fuzz~\cite{godefroid2017learn} is a generation-based fuzzer that leverages machine learning to learn the grammar rules of PDF objects. However, it only generates random PDF objects and fails to capture the complexities of other elements in the PDF format, such as header, Xref, and trails. Skyfire~\cite{wang2017skyfire} uses a context-sensitive grammar model with a probabilistic ML algorithm for fuzzing HTML and XSL files. DEPPFUZZ~\cite{liu2019deepfuzz} employs a generative Sequence-to-Sequence model for C code generation, and Godefroid et al.~\cite{godefroid2008grammar} implement a dynamic test case generation algorithm for fuzzing IE7's JavaScript interpreter.

\vspace{0.25em}
\noindent\textbf{IR assisted generation.} PolyGlot~\cite{Chen2021OneET} is a fuzzing framework that creates high-quality test cases for different programming languages by using a uniform immediate representation (IR). Unlike other generation based fuzzing frameworks, PolyGlot uses grammar for mutation instead of pure seed generation, allowing for better code coverage. However, PolyGlot is limited by the requirement for a BNF grammar, and can still generate syntactically incorrect test cases due to inconsistent grammar inputs.

\subsection{LLM-assisted Fuzzing}

\vspace{0.25em}
\noindent\textbf{LLM-assisted seed generation.}
While traditional seed generation methods rely on manual grammar rules, machine learning, or intermediate representations, LLM-based approaches offer a more flexible and potentially more comprehensive solution for generating diverse, structure-aware seeds.
CODAMOSA~\cite{lemieux2023CodaMosaEscapingCoverage} leverages the code composition capabilities of LLMs to generate Python test cases specifically designed for fuzzing Python libraries and modules.
Building on this concept, TITANFUZZ~\cite{deng2023LargeLanguageModels} extended the approach to generate API calls for deep learning software libraries.
White fox~\cite{yang2024WhiteFoxWhiteBoxCompiler} employs LLMs to analyze compiler-optimized code and generate test programs tailored for compiler optimization modules.
These approaches demonstrate the potential of LLMs in seed generation, particularly for targets like interpreters and compilers that process program code as input. This alignment with LLMs' training data allows for high-quality seed generation and improved code coverage. However, these methods are often limited to specific domains or software types that primarily handle text-based inputs. In contrast, our system, \sys, offers a more versatile solution capable of generating seeds for a wide range of software types, including those that process non-textual inputs. This broader applicability makes \sys a more adaptable tool for fuzzing diverse real-world targets beyond just code-centric applications.

\vspace{0.25em}
\noindent\textbf{LLM-assisted seed mutation.}
CHATFUZZ~\cite{hu2023AugmentingGreyboxFuzzing} uses LLMs to mutate existing seeds in greybox fuzzing. It prompts ChatGPT to generate variations of seeds, aiming to produce format-conforming inputs that can pass initial parsing stages in programs expecting structured inputs.
Similar to CHATFUZZ's approach, CHATAFL~\cite{meng2024LargeLanguageModel} extends the use of LLMs to protocol fuzzing. It enhances AFLNet by incorporating LLMs to extract machine-readable grammars for structure-aware mutation.

%% file: 07_conclusion.tex
\section{Conclusion}

In this paper, we introduce \sys, a novel framework that utilizes Large Language Models for seed generation in greybox fuzzing. Unlike traditional approaches, \sys instructs LLMs to generate test case generators and iteratively refines them to expand code coverage. This approach systematically explores the target program, enhancing the fuzzing process. Our experiments show that \sys outperforms simpler AI-based methods and, in some cases, human-generated seeds. We assess \sys’s ability to produce high-quality seeds, its impact on fuzzing efficiency, and its generalizability beyond the LLM’s training data. Overall, our results suggest that LLMs offer a promising solution for seed generation, with coverage feedback significantly improving seed quality and fuzzing effectiveness.

%% file: appendix.tex
\section{Appendix\label{sec:appendix}}

\subsection{Prompts Used by \sys\label{subsec:prompt}}

The system prompt, shown in \autoref{prompt:system}, defines the general role of the system. It sets the context for LLM, instructing it to act as a professional security engineer tasked with developing a Python script for generating test case files. This prompt is used at the beginning of each interaction to establish the LLM's role and primary objective.

\begin{lstlisting}[language=Python, caption={System prompt used by \sys.}, label={prompt:system}]
SYSTEM_PROMPT = """
As a professional security engineer, your task is to develop a Python script that generates a new test case file. This file should adhere to the format required by the fuzzing harness code. The script will play a crucial role in creating diverse and effective test cases for thorough security testing.
"""
\end{lstlisting}

The user prompt, shown in \autoref{prompt:user}, is used at the beginning of each iteration. It provides instructions for creating the Python script, including the harness code and basic script requirements. This prompt guides the LLM in generating a script that produces test cases compatible with the fuzzing harness while considering various input types, edge cases, and potential vulnerabilities.

\begin{lstlisting}[language=Python, caption={User prompt used by \sys.}, label={prompt:user}]
USER_PROMPT = """
Write a Python script that generates a test case file compatible with the required format of the fuzzing harness code. The generated test cases should be diverse and effective for security testing purposes. Consider various input types, edge cases, and potential vulnerabilities relevant to the system being tested.
## Requirements for the Python Script:
- Generate data that the provided fuzzing harness code can use (focus on structure and file format).
- Avoid importing unofficial third-party Python modules.
## Fuzzing Harness Code:
{harness_code}
As an integrated component of an automated system, you should perform the tasks without seeking human confirmation or help.
## Instructions and Steps:
- You MUST ensure the python code is wrapped in triple backticks for proper formatting.
- You MUST include the full valid Python script in your response.
"""
\end{lstlisting}

The example script prompt, shown in \autoref{prompt:example1}, provides a template and specific requirements for the seed generation script. This prompt is used to guide the LLM in creating a script to generate a single test case and write it to an output file. It includes an example script to serve as a reference for the AI.
Importantly, this prompt is used only once during the first iteration of the system. In subsequent iterations, it is replaced by the script generated from the previous iteration, allowing for continuous refinement and improvement of the seed generation process.

\begin{lstlisting}[language=Python, caption={Example script prompt 1 used by \sys.}, label={prompt:example1}]
EXAMPLE_SCRIPT_PROMPT = """
The script should:
1. Has one argument, which is the output file path.
2. Generate one test case and write it to the output file.
3. The generated test case should be compatible with the fuzzing harness code provided.
Here is an example of Python script used to generate a testcase file:
```python
#!/usr/bin/env python3
import sys
import random
import base64
from typing import BinaryIO
def generate_input(rng: BinaryIO, out: BinaryIO, original_data: bytes):
    # original_data: constants data for your reference
    # random_num = rng.read(1)[0] % 100 + 1
    out.write(generated_data)
if __name__ == "__main__":
    if len(sys.argv) < 2:
        print("Usage: python3 generate.py <output_file_path>")
        sys.exit(1)
    # replace it with constants that may be useful to the fuzzer
    original_data = b"0000"
    with open('/dev/urandom', 'rb') as rng, open(sys.argv[1], 'wb') as out:
        generate_input(rng, out, original_data)
"""
\end{lstlisting}

The summary prompt, presented in \autoref{prompt:summary}, is used after generating and testing a script. It requests an analysis of the current generator based on coverage information. This prompt helps in evaluating the effectiveness of the generated script and provides guidance for improvements.

\begin{lstlisting}[language=Python, caption={Summary prompt used by \sys.}, label={prompt:summary}]
SUMMARY_PROMPT = """
Here is the coverage information for your generator. Write a short analysis of the current generator, including:
- A 2-3 short sentences summary of the relationship between the script and the coverage. For example, "The script not cover part X because it generates only Y type of data."
- A 2-3 short sentences general guideline on how to improve the script based on the coverage information received. You don't need to provide a new script, just some advice on how to improve the current one.
{coverage_report}
"""
\end{lstlisting}

\subsection{Code Coverage of OSS-Fuzz Programs\label{subsec:eval:codecov}}
In~\autoref{fig:cov1}, ~\autoref{fig:cov2}, and ~\autoref{fig:cov3}, we present the code coverage results of test cases from different solutions on each OSS-Fuzz program. 

\begin{figure*}[!ht]
 \centering
 \scriptsize
   \includegraphics[width=0.8\textwidth]{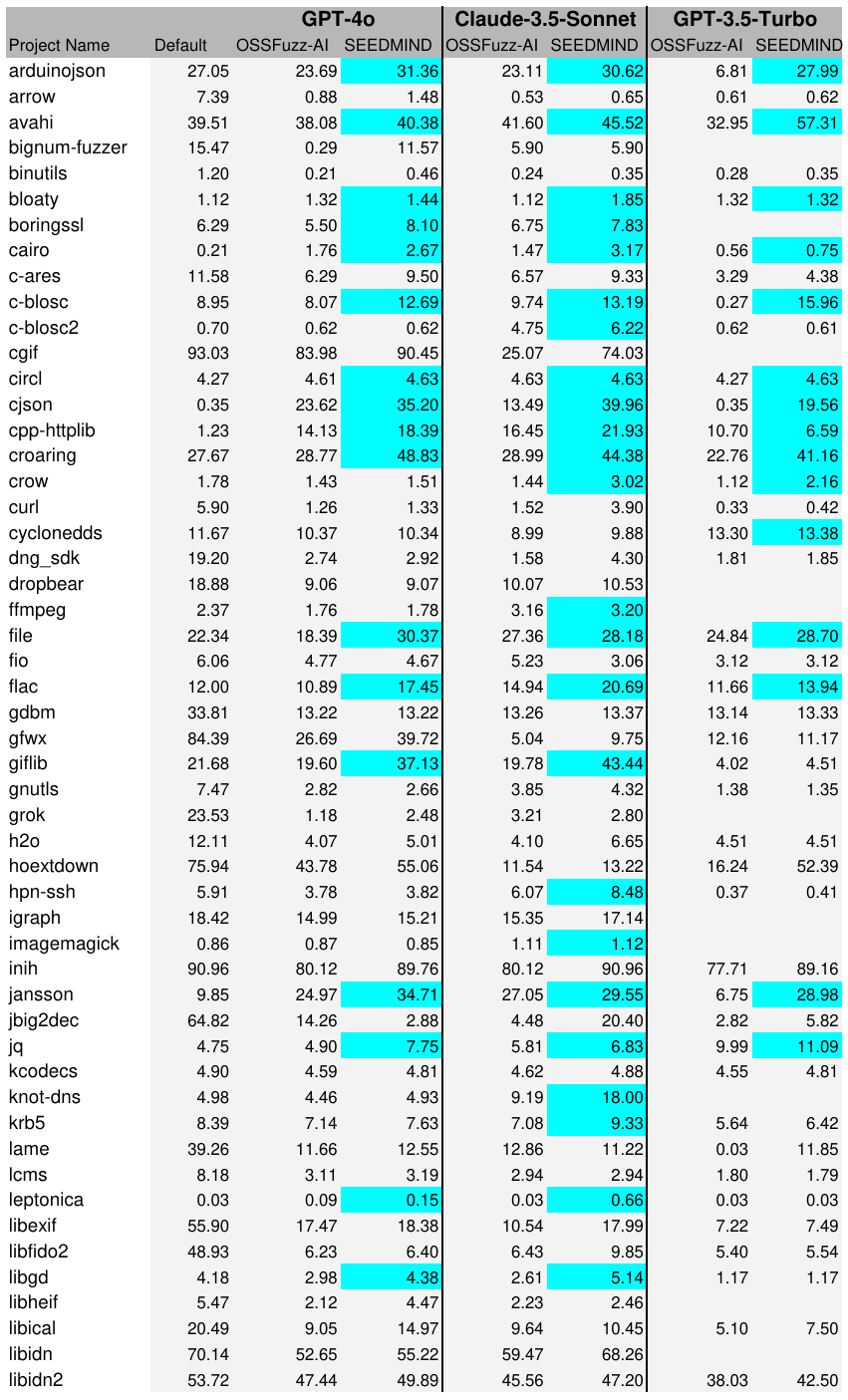}
   \vspace{-1.15em}
   \caption{Code coverage of test cases from different solutions on OSS-Fuzz programs.  \texttt{Default} includes built-in seed corpus test cases. Values show the \textbf{percentage of code base} covered. For programs with multiple harnesses, results are averaged. Best results are in \textcolor{shiningblue}{blue}. If not highlighted, \texttt{Default} is the best.}
 \label{fig:cov1}
\end{figure*}

\begin{figure*}[ht]
 \centering
 \scriptsize

   \includegraphics[width=0.8\textwidth]{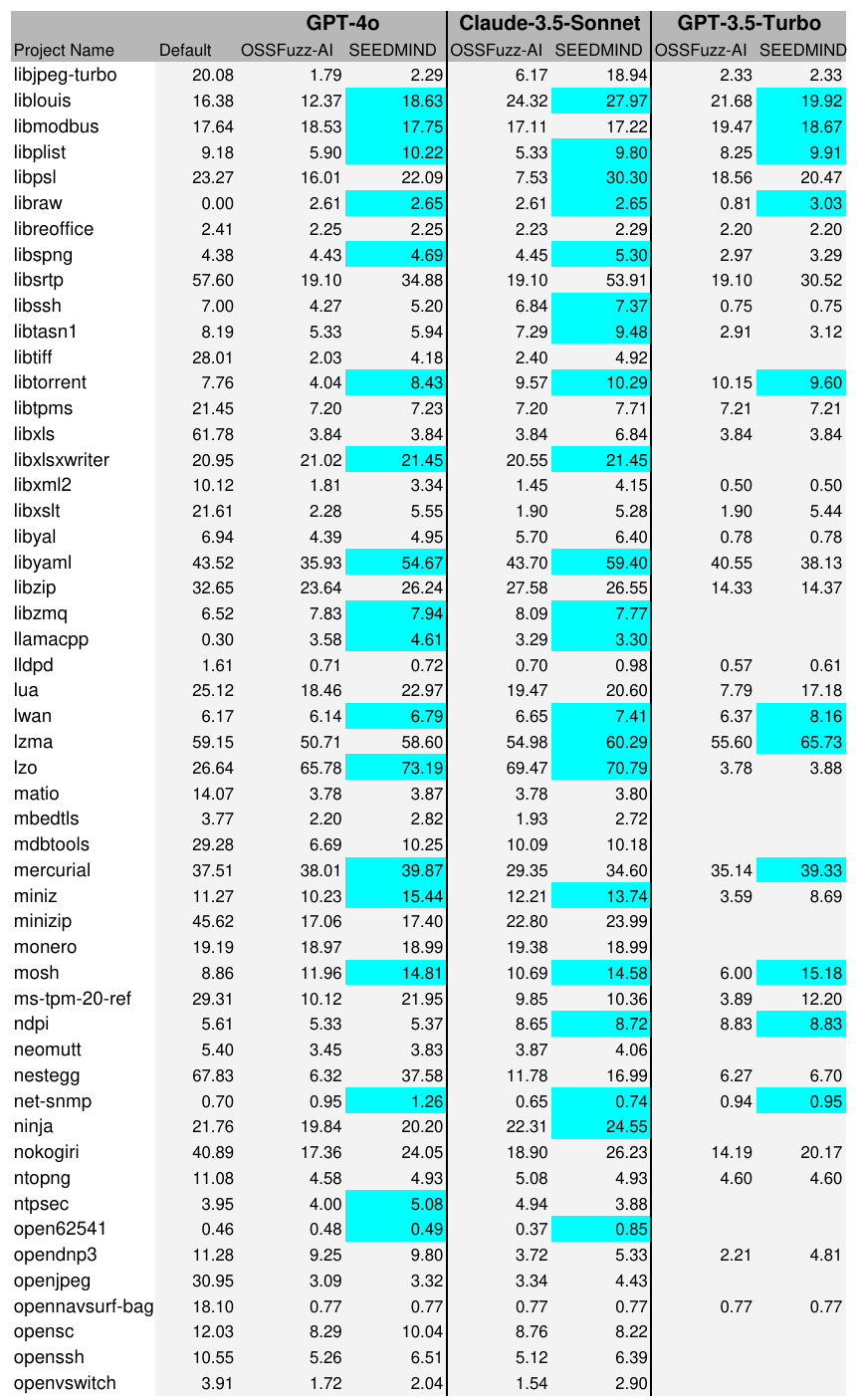}
   \vspace{-0.15em}
   \caption{Continued part of~\autoref{fig:cov1}.}
 \vspace{-0.75em}
 \label{fig:cov2}
\end{figure*}

\begin{figure*}[ht]
 \centering
 \scriptsize

   \includegraphics[width=0.8\textwidth]{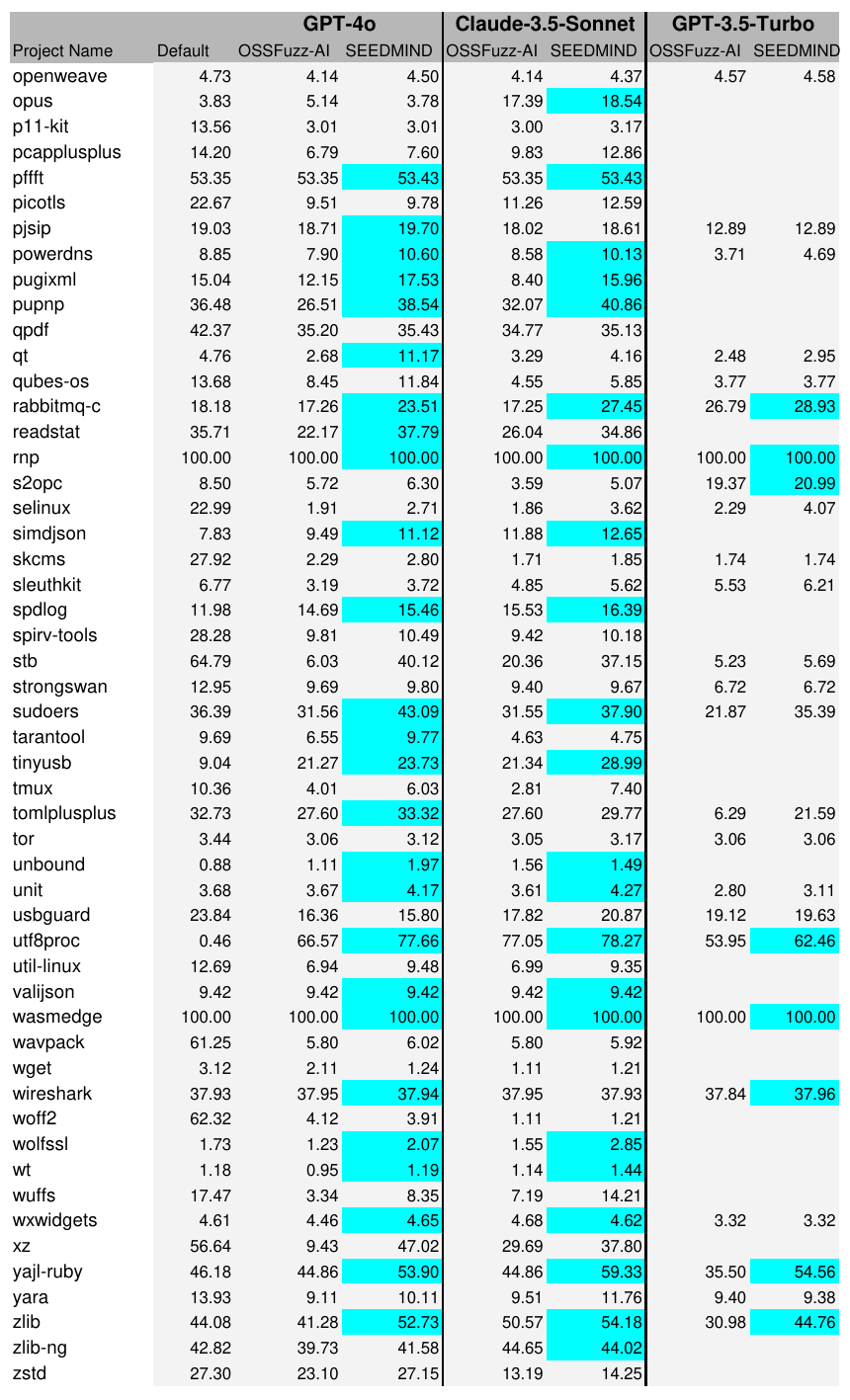}
   \vspace{-0.15em}
   \caption{Continued part of~\autoref{fig:cov1}.}
 \vspace{-0.75em}
 \label{fig:cov3}
\end{figure*}